# Beamsteering on Mobile Devices: Network Capacity and Client Efficiency

Hang Yu, Lin Zhong, and Ashutosh Sabharwal
Department of Electrical & Computer Engineering, Rice University, Houston, TX 77005
{Hang.Yu, lzhong, ashu}@rice.edu

**ABSTRACT**

Current and emerging mobile devices are omni directional in wireless communication. Such omni directionality not only limits device energy efficiency but also poses a significant challenge toward the capacity of wireless networks through inter-link interference. In this work, we seek to make mobile clients directional with beamsteering. We first demonstrate that beamsteering is already feasible to mobile devices such as Netbooks and eBook readers in terms of form factor, power efficiency, and device mobility. We further reveal that beamsteering mobile clients face a unique challenge to balance client efficiency and network capacity. There is an optimal operating point for a beamsteering mobile client in terms of the number of antennas and transmit power that achieve the required capacity with lowest power. Finally, we provide a distributed algorithm called BeamAdapt that allows each client to closely approach its optimal point iteratively without central coordination. We also offer a cellular system realization of BeamAdapt. Using Qualnet-based simulation, we show that BeamAdapt with four antennas can reduce client power consumption by 55% while maintaining a required network throughput for a large-scale network, compared to the same network with omni directional mobile clients.

## 1. Introduction

All existing and emerging wireless standards assume that their mobile clients are omni directional, radiating power toward all directions. Such omni directionality has become a critical barrier to not only network capacity but also the client efficiency as the number of mobile clients explodes, thanks to the popularity of smartphones, NetBooks, and eBook readers such as Kindle and upcoming Apple iPad.

In this work, we study a client-based approach toward addressing the omni directionality: using beamsteering on mobile devices for directional transmission. By focusing the transmit power toward the right direction, beamsteering can not only improve SNR at the intended receiver but also reduce the interference to peer links. While beamsteering has been studied and deployed for base stations, access points, and even vehicles, it has never been examined for mobile devices due to its large form factor and power overhead, as exemplified by the Phocus Array systems used in many recent works, e.g. [1-3].

Therefore, we first show that the form factor and power overhead of beamsteering is not fundamental. Using historical data of industrial designs, we demonstrate that beamsteering is not only feasible but also efficient for mobile devices. Beamsteering with four antennas can already fit on mobile devices such as Kindle and iPad and its circuit power overhead can be more than compensated by reduced transmit power. We then experimentally demonstrate that the beamsteering gain remains high even when the client can not only move but also rotate, even under indoor non-line-of-sight (NLOS) environments.

Furthermore, we analyze two tradeoffs made by uplink beamsteering using a dynamic number of antennas. 1) Beamsteering makes tradeoffs between circuit and transmit power. By using more antennas, it incurs higher circuit power but forms a more directed beam with higher gain. As a result, lower transmit power is required to deliver the same received signal strength. 2) Beamsteering makes tradeoffs between network capacity and client efficiency. The number of antennas that gives the highest client efficiency is not necessarily the one with least interference to peers. Because of such tradeoffs, there is an optimal operating point for each client in terms of the number of antennas and transmit power to deliver the same uplink capacity with the lowest power consumption.

Finally, to approach the optimal operating points for mobile clients of a large-scale network, we provide a distributed algorithm, called *BeamAdapt*, with which each mobile client iteratively adjusts its number of antennas and transmit power without centralized coordination. We show that BeamAdapt has guaranteed convergence and its solution approaches the optimal despite that the problem is non-convex. We further offer a system design of BeamAdapt in the context of modern cellular networks. We evaluate the system realization of BeamAdapt with Qualnet-based simulation of a large-scale network. The results show that averagely BeamAdapt can reduce client power consumption by 40% and 55% with two and four antennas, respectively, while maintaining the same network throughput.

In summary, we make the following contributions toward beamsteering on mobile devices.

- Through both analysis and experimentation, we demonstrate that beamsteering is not only feasible but also efficient for form-factor and power-constrained mobile devices such as netbook, Kindle and iPad.

- We analyze the tradeoffs between network capacity and client efficiency made possible by beamsteering mobile clients and show that beamsteering mobile clients can improve the network capacity and client efficiency over conventional omni directional clients.
- We offer BeamAdapt, a distributed algorithm that provides close-to-optimal solutions with guaranteed convergence. We show that BeamAdapt can be efficiently realized in large-scale cellular systems and provides significant improvement in client efficiency.

Making mobile clients directional is a radical departure from existing and emerging wireless technologies. While we demonstrate the potential benefits from beamsteering clients in network capacity and client efficiency, much more research is needed in all layers of the network system in order to fully realize such potential.

The rest of the paper is organized as follows. Section 2 presents background knowledge on beamforming and beamsteering. Section 3 presents a feasibility study of using beamsteering on mobile devices. Section 4 illustrates and analyzes the two tradeoffs made by beamsteering. Section 5 provides the theoretical framework of BeamAdapt and Section 6 presents its system design within modern cellular network systems. Section 7 evaluates BeamAdapt with Qualnet simulation. Section 8 addresses related works and Section 9 concludes the paper.

## 2. Beamforming Primer

We first provide a brief overview of the general beamforming technology and its special form, beamsteering.

### 2.1 Beamforming

Beamforming uses a group of antennas to transmit and receive radio signals. Each antenna has a devoted RF chain that bridges the baseband signal and RF signal. It forms a beam pattern by assigning a complex weight to each baseband signal and combining them to form the array output, or

$$y(t) = \mathbf{w} \cdot \mathbf{x}(t) = \sum_{n=1}^{N} w_n x_n(t)$$

where the baseband signal vector, weight vector and array output are denoted as $\mathbf{x}(t)$, $\mathbf{w}$ and $y(t)$, respectively. Note that for transmit beamforming, $\mathbf{x}(t)$ has identical elements. The beamforming gain $G$ is defined as the ratio of the received signal power by beamforming to that by an omni-directional pattern. Notably, $G$ is dependent on the signal departure direction $\theta$ as well as the number of antennas, or beamforming size $N$, or $G=G(N,\theta)$. The direction with maximum gain, $G_{max}$, is called *look direction*.

There are two types of beamforming techniques: one assumes no interference knowledge and optimizes the beam pattern for the intended receiver only, named *beamsteering*; the other one intentionally suppresses interference to unintended receivers, called *adaptive beamforming*. In this work, we choose to use beamsteering on mobile devices for the following reasons [4]. 1) the advantage of adaptive beamforming over beamsteering in terms of interference reduction is very limited for mobile clients because of the small number of transmit antennas and relatively large number of interfered receivers in mobile networks. 2) adaptive beamforming requires accurate interference knowledge, which is expensive to acquire in a mobile network. 3) a mobile client, especially in self-managed networks, is usually reluctant to offer benefit to other clients by sacrificing its own link performance.

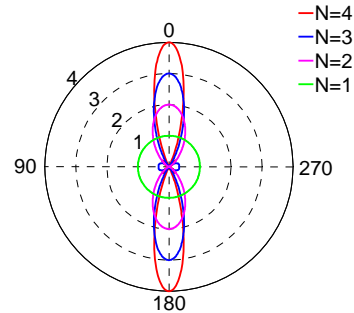

**Figure 1. Beamsteering pattern of a linear array with different beamsteering sizes ($N$)**

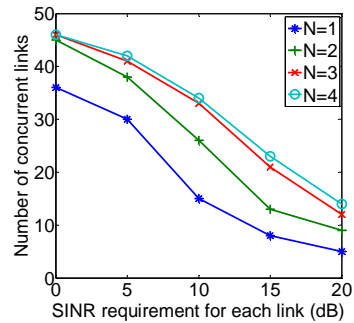

**Figure 2. Number of concurrent links allowed by beamsteering with different sizes ($N$)**

Although beamforming is mainly considered effective outdoor, recent work [3, 5] has shown it can achieve good gain indoor too.

### 2.2 Beamsteering

Beamsteering is a special form of beamforming. In beamsteering, the weight vector is assigned as $\mathbf{w}=\mathbf{h}^*$, where $\mathbf{h}$ is the channel vector with each of its elements representing the channel coefficient between a transmit antenna and the receiver. Therefore, given $\mathbf{h}$, the weight vector is also given without any additional computation or signal processing. More generally, the channel vector $\mathbf{h}$ is denoted as *channel state information* (CSI). For transmit beamsteering, CSI can only be estimated. There are two ways of CSI estimation: explicit and implicit. For explicit CSI estimation, the receiver leverages the fixed training symbols sent from the transmitter to calculate the channel coefficients and send it back to the transmitter. This procedure is repeated for each transmit antenna. For implicit CSI estimation, the transmitter estimates the reverse channel when receiving and uses it for transmitting. We note that implicit CSI estimation requires channel reciprocity to be effective.

With perfect CSI, beamsteering can achieve the maximum gain in the look direction while less gain in others. In the look direction, signals from each transmit antenna will add coherently at the receiver. Therefore, the maximum gain, $G_{max}$, is equal to the beamsteering size $N$ [4]. Figure 1 shows the beam patterns for beamsteering sizes from one to four.

The network capacity improvement of beamsteering has been widely appreciated. To generally illustrate the benefit of using beamsteering in mobile networks, Figure 2 shows the maximum number of links that are allowed to be active simultaneously (*feasible set*) under beamsteering with different sizes and an omni-directional beam pattern. Clearly, beamsteering enables more active links by reducing inter-link interference. The larger the beamsteering size, the larger the number of concurrent links. In addition, for each SINR requirement, same amount of power has been employed by beamsteering with different sizes. Therefore, the capacity improvement is on the basis of no extra power expense.

## 3. Feasibility Check

Few have considered beamsteering for use on mobile devices. Beamsteering systems were known to be bulky and power-hungry, e.g. the Phocus Array system used in many recent works, e.g. [1-3]. Moreover, as a mobile device can not only move but also rotate, CSI estimation, especially the phase estimation of the channel coefficients, can be challenging. Therefore, the first question we seek to answer is: *is beamsteering feasible on a mobile device?* We next examine three conventional constraints known as technical barriers: form factor, power efficiency, and device mobility.

### 3.1 Form Factor

Beamforming systems have conventionally been bulky due to not only the size of multiple integrated RF circuits, but also the antenna spacing requirement. However, we note beamsteering systems can fit into a mobile device for the following reasons. 1) Technological advancement has reduced the size of integrated circuits significantly. Single-chip transceivers integrating multiple RF chains are now feasible, e.g. [6]. 2) The antenna spacing constraint can be relaxed for beamsteering. Unlike MIMO systems, beamsteering allows antenna spacing to be as small as 0.5$\lambda$. This already makes it possible for a 2GHz laptop, Netbook or E-Book reader such as Amazon Kindle or Apple iPad, to have a linear array with four antennas. For even smaller mobile devices such as Smartphones, a circular array can fit. Moreover, as radio moves to higher frequency bands such as 60GHz, the antenna spacing constraint will vanish.

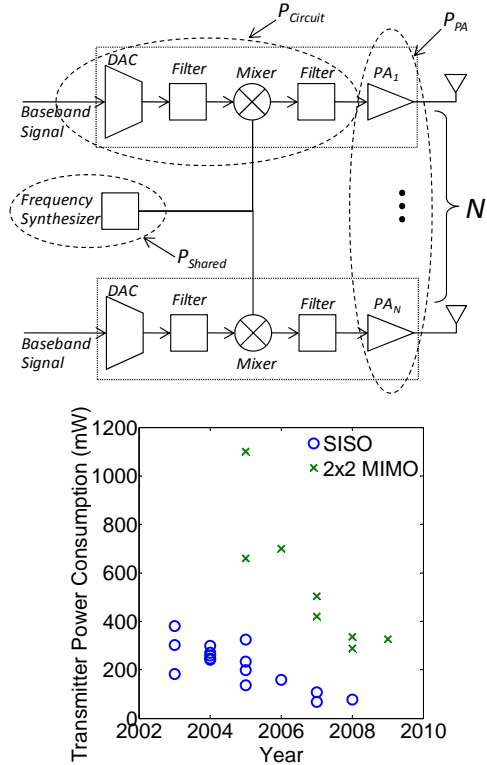

**Figure 3.** (Top) RF components of beamsteering transmitter; (Bottom) transmitter power trend from designs reported in ISSCC and JSSC

### 3.2 Power Characteristics

By the use of multiple RF chains, beamsteering is also considered power-hungry. As indicated in Section 2, we claim that beamsteering incurs very little overhead in baseband processing. Therefore, we focus on the RF power characteristics only. To appreciate its power characteristics, Figure 3 illustrates the major hardware components of a beamsteering transmitter. We can see that the transmitter power can be decomposed into that of the circuitry shared by all active RF chains, e.g. the frequency synthesizer, and that of each RF chain. Let $P_{Shared}$ denote the power contributed by the shared circuitry. The power contributed by each active RF chain can be broken down to that by the power amplifier, $P_{PA}$, and that by the rest of RF circuitry, $P_{Circuit}$. We approximate $P_{PA}$ as a linear function of the transmit power, $P_{PA}=(1+\alpha)P_{TX}$. While a linear power model is not absolutely accurate due to possibly different modulation schemes and the intrinsic nonlinearity of power amplifier efficiency, it is considered a good approximation [7]. Therefore we can estimate the total power $P$ as

$$P=(1+\alpha)P_{TX}+NP_{Circuit}+P_{Shared} \qquad (1)$$

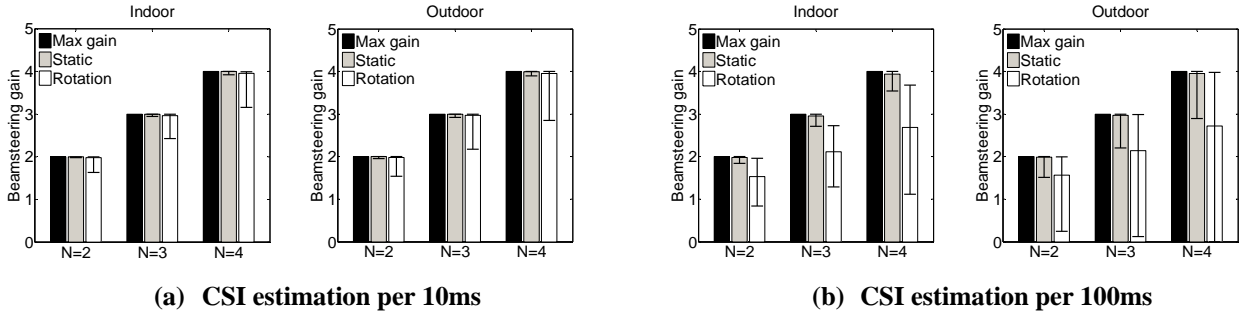

(a) CSI estimation per 10ms      (b) CSI estimation per 100ms

**Figure 4. Beamsteering gain under CSI estimation. For those from measurement, we also report the range of beamsteering gain.**

In the rest of the paper, we adopt parameters of Equations (1) according to [7]. That is, $\alpha$=1.875, $P_{Circuit}$=48.2mW, and $P_{Shared}$=50mW. They are on par with state-of-the-art commercial transceivers in 2-5GHz band [8, 9].

### 3.2.1 Power Trend of CMOS Transceivers

Although RF integrated circuits improve slower than their digital counterparts, their power efficiency still follows Moore's Law. To illustrate this trend, we have examined the CMOS transceiver realizations reported in ISSCC [10] and JSSC [11], the top conference and journal for semiconductor circuits, from 2003 to 2009, and show their circuit power consumption, $P_{Circuit}+P_{Shared}$, in Figure 3. The figure clearly shows the continuous improvement in the power efficiency of both SISO and MIMO transceivers. For a concrete example, a transceiver with the 0.18$\mu$m CMOS technology consumes 270mW in transmit [12], while a similar one implemented with the 65nm CMOS technology merely consumes 35mW [13]. As semiconductor process technologies continue to improve according to Moore's Law, $P_{Circuit}$ and $P_{Shared}$ will continue decreasing. A beamsteering system with four RF chains is very likely to consume less than 50mW for the RF circuitry in the near future and $P_{PA}$ will increasingly dominate the total transmitter power consumption. By focusing transmit power toward the intended direction, beamsteering systems can indeed reduce $P_{TX}$ and therefore actually improve the transceiver power efficiency, as we will further show in Section 4.

### 3.3 CSI Estimation under Device Mobility

The final challenge to beamsteering from a mobile device is that a mobile device can not only move but also rotate. Recent work has shown that beamforming can cope with even vehicular mobility very well, e.g. [1, 2]. However, since beamsteering is sensitive to the phase change of the channel coefficient, device rotation can potentially introduce even faster channel fading. We next experimentally evaluate the gain of beamsteering under device rotation.

We perform the experiments with two WARP nodes from Mango Wireless [14]. We build a beamsteering array with four antennas 0.5$\lambda$ from each other on one node as a mobile client. The client and AP nodes are placed as far as the WARP nodes allow, about 15 meters in our experiments. In the experiments, the beamsteering mobile client continuously sends frames with training symbols to the other node every 10ms and the latter sends back the estimated CSI through an Ethernet cable. Therefore, the mobile client updates the CSI each 10ms, calculates the weight vector and forms a beam accordingly. To challenge the CSI estimation, we rotate the mobile client with a computerized motor at 180°/s. We repeat the experiments both indoor and outdoor. The indoor environment is in an office building with active network usage at 2.4GHz and without line-of-sight paths.

While we could not simultaneously examine different beamsteering sizes and different CSI estimation frequencies in real time, we have collected traces of the channel coefficients which allow us to emulate the channel offline. That is, we replay the channel using the recorded traces but assume different beamsteering sizes ($N$=2,3,4), and different CSI estimation frequencies (10ms and 100ms). Since the average beamsteering gain is only dependent on the CSI, the offline emulation gives identical results as real-time evaluation does.

The key question we aim to answer from experiments is: what is the impact of device rotation on CSI estimation and beamsteering gain? To see this, Figure 4 shows the average beamsteering gain under CSI estimation with and without device rotation. In each sub-figure of Figure 4, three values of beamsteering gain for each beamsteering size ($N$=2,3,4) are shown: the upper bound given by perfect CSI, the one given by estimated CSI with stationary client, the one given by estimated CSI with rotating client. One can see that when the CSI estimation interval is 10ms, the CSI can be very accurate even with client rotational speed of 180°/s. When the interval is increased to 100ms, the beamsteering gain will be affected by client rotation. The rotation has higher impact for larger beamforming sizes with potentially more focused beams. Therefore, we conclude that under high speed device rotation such as 180°/s, beamsteering can be effective with reasonable CSI estimation intervals, e.g., 10ms. Meanwhile, it can potentially become less ef-fective with longer CSI estimation intervals, e.g., 100ms. This indicates one should intelligently use beamsteering to circumvent the performance drop by out-of-date CSI. In addition, we observe that the performance of CSI estimation is more stable indoor, due to the rich multipath effect. This can be seen from the range of beamsteering gain in each sub-figure.

**Table 1. Settings for the one-link and two-link network simulation**

| Parameters | Values |
|---|---|
| d (distance) | 0.5km |
| Max beamsteering size | 4 |
| Transmit power decay factor | 4 |
| Receiver thermal noise | -170dBm/Hz |
| Channel bandwidth | 5MHz |
| Carrier frequency | 2GHz |

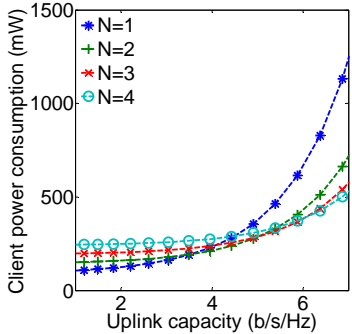

**Figure 5. Client power consumption to deliver a given link capacity**

## 4. Tradeoff Analysis of Beamsteering

We next analyze two important tradeoffs made by beamsteering for a mobile client and the network of mobile clients.

### 4.1 Circuit Power vs. Transmit Power

Compared to omni directional antennas, beamsteering essentially reduces transmit power $P_{TX}$ and increases circuit power $P_{Circuit}$. As the circuit is increasingly efficient, the tradeoff made by beamsteering is increasingly profitable. We illustrate this by analyzing an uplink channel between a mobile client and its base station. We assume a line-of-sight (LOS) propagation and adopt the parameters as specified in Table 1. Figure 5 shows the client power consumption calculated by Equation (1) to deliver a range of channel capacity for beamsteering sizes from one to four with the power parameters in Section 3.2.

One can make the following conclusions from the figure.

- First, beamsteering ($N>1$) is already more efficient than omni-directional transmission when delivering a capacity of 3.6b/s/Hz or higher.

- Second, the larger the required link capacity, the larger the most efficient beamsteering size. This shows that beamsteering is actually increasingly efficient in delivering higher capacity.

- Finally, given the transmit power decay factor and distance, one can derive the required transmit power for omni directional transmission $P_O$, to achieve certain link capacity. Recall that the beamsteering gain $G_{max}$ is equal to the beamsteering size, $N$. Therefore, a transmit power of $P_O/N$ is needed. Using Equation (1), we can compute the most efficient beamsteering size as $N_{opt} = \sqrt{(1+\alpha)P_O/P_{Circuit}}$. Apparently, beamsteering is increasingly more efficient as $P_{Circuit}$ decreases (Section 3.2) and $\alpha$ increases for more advanced modulation methods [7].

The research question is: *given the link capacity requirement, how to find out the most energy efficient tradeoff between circuit and transmit power?* While the answer may appear to be simple with this single-link example, we next show it can be very challenging to obtain in a network of mobile clients when inter-link interference is considered.

### 4.2 Network Capacity vs. Client Efficiency

Beamsteering can significantly reduce network interference by focusing radiation energy to the intended receiver and reducing it to the others. In general, the larger the beamsteering size, the less the interference. However, as shown above, the most efficient beamsteering size is determined by the required link capacity. As a result, beamsteering must balance peer interference and link power consumption, or between network capacity and client efficiency.

We examine such tradeoffs by adding another link to the single-link example analyzed in previous section, illustrated by Figure 6. Again, we focus on uplinks and adopt the same simulation settings in Table 1. For simplicity, we assume the two links are symmetric, i.e. $d_{11}=d_{22}$ and $d_{12}=d_{21}$. We adopt the aggregate interference model in [15] to calculate the SINR. The network power consumption, $P_{Network}$, is calculated as the total power consumption by the two mobile clients and the network capacity, $C_{Network}$, is defined as the aggregated capacity of the two uplinks or $C_{Network}=C_1+C_2$.

Figure 7 shows the tradeoffs between $P_{Network}$ and $C_{Network}$ by beamsteering with one to four antennas under various network properties. Data in each subfigure is calculated for a given $d_{11}/d_{22}$ (2 or 5) and $C_1/C_2$ (2 or 5). Each subfigure shows five plots. The first four plots are generated as the mobile clients have same beamsteering sizes, from one to four ($N=1$ to 4). The fifth plot is generated as the two mobile clients are allowed to have different beamsteering sizes that lead to the lowest $P_{Network}$ for the given $C_{Network}$. Therefore, it is the optimal case that yields the best tradeoff between network capacity and client efficiency (*Opt-BeamSteering*).

We make several observations from Figure 7.

- Beamsteering ($N>1$) is more efficient to deliver much of the network capacity. This is more evident than the single-link case in Figure 5 because of the addition of interference between two links.
- With same power, beamsteering ($N>1$) achieves higher network capacity than non-beamsteering ($N=1$). This attests to the interference reduction by beamsteering.
- Beamsteering can achieve network capacity that is unattainable by non-beamsteering even if the latter uses unlimited power (the upper two figures in Figure 7). This further highlights the network capacity benefit of beamsteering from suppressing inter-link interference.
- The optimal beamsteering sizes given the network capacity are dependent on not only the physical properties $d_{11}/d_{22}$ but also the link capacity requirements $C_1/C_2$.
- Most importantly, the optimal beamsteering sizes of two clients need to be jointly decided, with inter-client coordination.

The question is: *how could mobile clients of a large network identify their beamsteering sizes that collectively minimize the network power consumption, without centralized coordination?* We embark on this question next.

## 5. BeamAdapt: Distributed Optimization of Beamsteering Mobile Clients

To answer the above research question, we next provide its theoretical formulation and a distributed algorithm, BeamAdapt.

### 5.1 Problem Formulation

As stated in Section 4.2, the optimal tradeoff between network capacity and client efficiency is given by the minimum network power consumption that achieves certain network capacity. That is, our objective is

*Minimize*

$$P_{Network} = \sum P_i, \forall 1 \leq i \leq M,$$

where the power of client $i$, $P_i$, is a function of the transmit power, $P_{TX,i}$, and the beamsteering size, $N_i$. Apparently $(P_{TX,i}, N_i)$ uniquely decides the beam pattern of client $i$.

Instead of posing a minimum constraint on $C_{Network}$, we individually constraining $C_i$, or equivalently the SINR of link $i$, $S_i$, such that $S_i \geq \rho_i$ where $\rho_i$ is the minimum SINR for base station $i$. The reason of separately constraining individual links is that different links usually have different capacity requirements. By appropriately tuning $\rho_i$, additional network properties such as individual link dissimilarity and fairness can be taken into account. We also consider the practical constraints on beamsteering size. That is, $N_i$ must be positive integers no greater than $N_{i,max}$, where $N_{i,max}$ is the number of antennas on client $i$. We allow different clients to have different $N_{i,max}$.

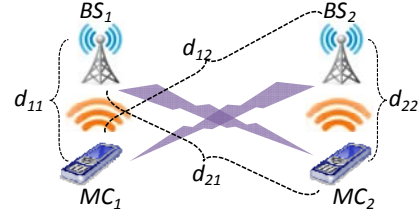

**Figure 6. Two-link network with two base stations and two mobile clients**

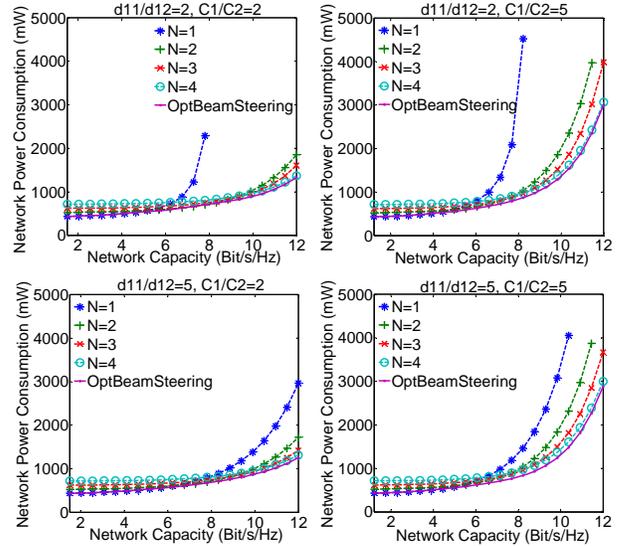

**Figure 7. Relation between network capacity $C_{Network}$ and network power consumption $P_{Network}$ for the two-link network**

To summarize, we formulate the optimization problem as follows:

*Minimize*

$$P_{Network} = \sum P_i(P_{TX,i}, N_i),$$

*s.t.*

$$S_i(\boldsymbol{P_{TX}}, \boldsymbol{N}) \geq \rho_i, 1 \leq N_i \leq N_{i,max}, \forall 1 \leq i \leq M,$$

where

$$\boldsymbol{P_{TX}} = (P_{TX,1}, \cdots, P_{TX,M}), \boldsymbol{N} = (N_1, \cdots, N_M).$$

We assume that $\rho_i$ is set properly so that the $M$ links compose a feasible set [15] to ensure that the problem has a solution. Nonetheless, while the problem is solvable, the constraints make finding the solution extremely hard. Firstly, each of the SINR constraints can be decided by all $2M$ optimization variables. The SINR function is non-convex with respect to these variables, which eventually yields the non-convexity of the problem. Secondly, there is no closed-form formulation of beamsteering gain $G$ as a function of $N$. Its dependence on angle $\theta$ makes low-order approximation hardly possible. Finally, the integer constraint of beamsteering sizes renders a NP-hard mixed integer programming (MIP) problem [16]. While an exhaustive searching algorithm can ultimately offer the solution, the complexity can be as high as $O\left(\prod_{i=1}^{M}(N_{i,max})\right)$, which makes it undesirable with large $M$. Most importantly, such exhaustive searching algorithm still requires all the clients in the network have knowledge of each other and cooperatively choose their beam patterns thereby does not offer answers to our research question.

To tackle this, we next introduce an iterative algorithm, BeamAdapt, which yields a close-to-optimal solution with much lower complexity and works in a distributed manner, not requiring each client to acquire knowledge of other links.

## 5.2 Distributed Algorithm: BeamAdapt

First we rewrite the problem as multiple sub-problems, i.e., the $i$th problem ($i = 1,2,\cdots M$) is

$$min_{P_{TX,i},N_i} P_i, s.t., S_i \geq \rho_i.$$

The optimal beam pattern $(P_{TX,i}, N_i)$ is adjusted iteratively. That is, we assume the transmit power and beamsteering size are $P_{TX}^{(k-1)}$ and $N^{(k-1)}$ for the $(k$-1$)$th iteration[1], and the measured link SINR is $S^{(k-1)}$, then for the $k$th iteration, $P_{TX}^{(k)}$ and $N^{(k)}$ can be obtained by solving the following optimization problem:

*Minimize*

$$(1 + \alpha)P_{TX}^{(k)} + N^{(k)}P_{Circuit} + P_{Shared}$$

*s.t.*

$$\frac{P_{TX}^{(k)}N^{(k)}}{P_{TX}^{(k-1)}N^{(k-1)}} = \frac{\rho}{S^{(k-1)}}, N^{(k)} \geq N^{(k-1)}.$$

The initial beamsteering size is set as one (omni directional pattern), i.e. $N^{(0)} = 1$, while $P_{TX}^{(0)}$ can be arbitrary.

The iteration stops when $|S^{(k)} - \rho| \leq \varepsilon$, where $\varepsilon$ can be set according to the requirement of accuracy, and then $(P_{TX}^{(k)}, N^{(k)})$ is accepted as the optimal beam pattern. In each iteration, $(P_{TX}^{(k)}, N^{(k)})$ can be obtained by a simple searching among all the feasible beam patterns, which has a complexity of $O\left(\max(N_{i,max})\right)$. Algorithm 1 shows the pseudo-code of the algorithm.

We note that when $M$=1, the problem reduces to single-link optimization with the tradeoff between circuit and transmit power. Since interference is absent thereby the received SINR is uniquely dependent on the transmit signal

[1] Here we have omitted the subscript $i$ for the notations since all clients employ the same algorithm.

**Algorithm 1: Pick the optimal beam pattern by BeamAdapt**

**Input**: set of concurrent links $\{1,2,\cdots,M\}$, SINR requirement $\{\rho_i, 1 \leq i \leq M\}$
**Output**: optimal beam pattern $\{(P_{TX,i}^{opt}, N_i^{opt}), 1 \leq i \leq M\}$

1  **for** $1 \leq i \leq M$
2   $k = 0, (P_{TX,i}^{(0)}, N_i^{(0)}) = (P_{TX,i}^{(0)}, 1), S_i = 0, P_{min} = +\infty$
3   $(P_{TX,i}^{opt}, N_i^{opt}) = (P_{TX,i}^{(0)}, N_i^{(0)})$
4   **while** $|S_i - \rho_i| \geq \varepsilon$
5    **for** $N_i^{(k)} \leq N_i^{(k+1)} \leq N_{i,max}$
6     Compute $P_{TX,i}^{(k+1)}, S_i$
7     $P_i = (1+\alpha)P_{TX,i}^{(k+1)} + N_i^{(k+1)}P_{Circuit} + P_{Shared}$
8     **if** $P_i \leq P_{min}$
9      $(P_{TX,i}^{opt}, N_i^{opt}) = (P_{TX,i}^{(k+1)}, N_i^{(k+1)})$
10     $P_{min} = P_i$
11    **end**
12   **end**
13   $k = k + 1$
14  **end**
15 **end**
16 **return** $\{(P_{TX,i}^{opt}, N_i^{opt}), 1 \leq i \leq M\}$

strength, one iteration can offer the solution which is guaranteed optimal.

## 5.3 Properties of BeamAdapt

We next evaluate BeamAdapt for its convergence and optimality.

### 5.3.1 Convergence

The iteration of BeamAdapt is guaranteed converged. We prove this by showing that solving the original problem is equivalent to solving finite sub-problems, each of which is isomorphic to a distributed power control problem that ensures convergence.

We divide the procedure of iteration into multiple stages, $k_l (1 \leq l \leq L)$, whereas in each stage the beamsteering size $N$ is constant throughout the iteration and only the transmit power $P_{TX}$ changes. In other words, the current stage $k_l$ evolves into $k_{l+1}$ when $N$ changes for at least one link. Based on the constraints $N^{(k)} \geq N^{(k-1)}$ we can easily get the following inequality

$$L \leq \prod_{i=1}^{M}(N_{i,max}) < +\infty,$$

which indicates finite $L$.

In each stage, the beamsteering size is fixed; therefore the original problem turns in to

*Minimize*

$$P_{Network} = \sum P_i(P_{TX,i}),$$

*s.t.*

$$S_i(\boldsymbol{P}_{TX}) \geq \rho_i, \forall 1 \leq i \leq M.$$

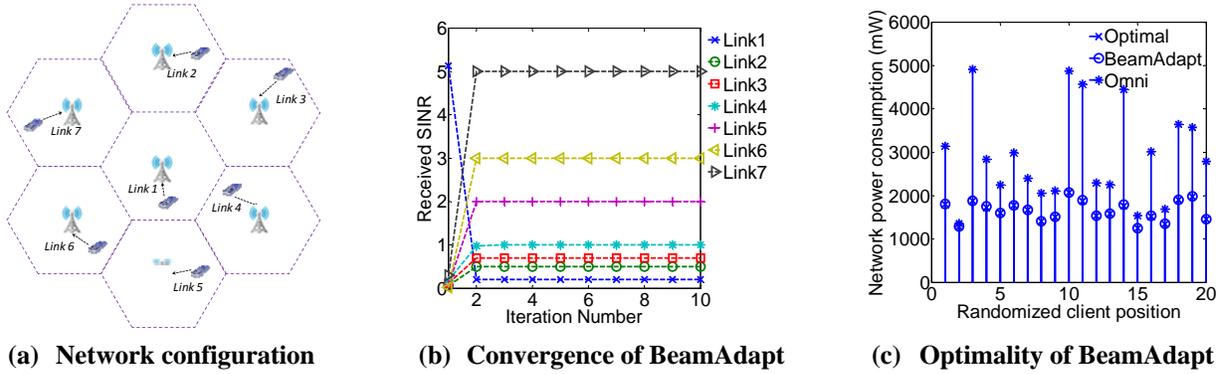

**(a) Network configuration**    **(b) Convergence of BeamAdapt**    **(c) Optimality of BeamAdapt**

**Figure 8. Convergence and optimality of BeamAdapt.**

It is isomorphic to a network power control problem where a distributed algorithm ensures convergence [17]. Therefore, by dividing the iteration procedure into finite stages each with guaranteed convergence, we can prove convergence of BeamAdapt.

We confirm the proof by using a network consisting of seven links illustrated in Figure 8(a). We set different SINR requires for the base stations. Figure 8(b) shows the convergence of the BeamAdapt. One can easily see that the convergence speed is fast, i.e., less than three iterations are needed to get very small $\varepsilon$.

*5.3.2 Optimality*

Next we evaluate how good the solution given by BeamAdapt. There is a chance that the BeamAdapt converges to a sub-optimal solution due to the non-convexity of the problem. However, we observe that the sub-optimal solution given by BeamAdapt is very close to optimal on average. For example, we compare the solution given by BeamAdapt with the optimal one using the same network in Figure 8(a). To eliminate the dependency of BeamAdapt on the network configuration, we randomize the position of each mobile client within its own cell. Assuming each client has four antennas or $N_{i,max}$=4, Figure 8(c) shows the solution given by BeamAdapt and the optimal solution in terms of network power consumption, for twenty randomized client positions. The figure also shows the solution with omni directional patterns for comparison. Clearly, the solution given by BeamAdapt is very close to the optimal and much better than that of the omni directional pattern: the network power consumption given by BeamAdapt is only 0.4% higher than that by the optimal solution on average.

## 6. Cellular System Realization of BeamAdapt

Next we elaborate the system design of BeamAdapt: we choose cellular network as an instructive scenario and implement BeamAdapt on the mobile access clients. BeamAdapt relieves clients in the network from acquiring network knowledge thereby avoids the overhead of information gathering or exchange. It also allows a client to use a heuristic method to shape the beam pattern thus automatically cope with channel variation and client mobility.

### 6.1 Cellular Networks

A cellular network is made up of a number of cells each served by a location-fixed base station which allows multiple clients to access. While the current cellular system (UMTS) is marked 3G, it is on the way to evolve into 4G, where 3GPP Long Term Evolution (LTE) [18] is known as the most promising candidate. In UMTS and LTE, channels between a UE (term for mobile clients) and a Node-B (term for base stations) are full-duplex, either by TDD or by FDD. The frequency reuse factor decides the frequency allocation to different cells. With frequency reuse factor of one, all cells share the same frequency bands and inter-cell interference occurs for each pair of them.

In cellular networks, uplink power control is employed for the UE. In both UMTS and LTE, uplink power control is performed in a closed-form manner. That is, the Node-B measures uplink SINR, decides the optimal transmit power and instructs the UE to set it as appropriate. The power control command from the Node-B is delivered to the UE through downlink control signaling.

### 6.2 Cellular Implementation Overview

We implement BeamAdapt in the PHY layer of the protocol hierarchy at the UE. The fundamental functionality of BeamAdapt is to identify and subsequently set the most energy efficient beam pattern. To cope with channel variation and uplink SINR requirement changes, BeamAdapt is performed in an adaptive manner. That is, the beam pattern is updated with the latest measurement of uplink SINR.

Figure 9 shows the basic structure of our BeamAdapt implementation. The main components include *CSI Estimation* and *Beam Pattern Adjustment*. The CSI is used to calculate the beamsteering weight vector. Beam pattern adjustment resides on top of uplink power control. Given the required transmit power, the most energy efficient beam pattern which delivers required signal strength to Node-B with least client power consumption can be identified.

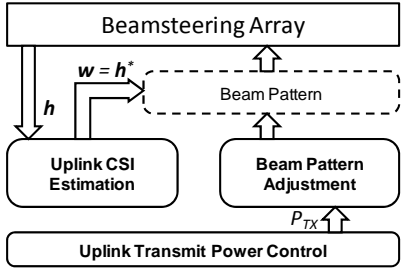

**Figure 10. BeamAdapt implementation in the PHY layer of the UE**

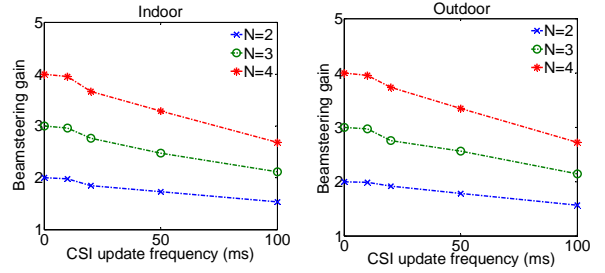

**Figure 9. Average beamsteering gain under different CSI update frequencies**

## 6.3 Key Components

Next we elaborate key components of the BeamAdapt implementation.

### 6.3.1 CSI Estimation

As elaborated in Section 2, implicit CSI estimation requires uplink and downlink channel reciprocity, which unfortunately, does not hold for long-range cellular channels, especially in FDD mode. Therefore, we adopt explicit CSI estimation in BeamAdapt. That is, the UE concatenates a short field made up of several training symbols to the data field in each uplink frame. Seeing the training symbols, the Node-B can estimate uplink CSI and feed it back to the UE during downlink control signaling and the UE forms a beam pattern as appropriate. The training field can be very short compared to the entire frame, i.e. a 16μs training field for beamsteering size of four and a 10ms frame [18]. The training symbols can be set in a similar way to MIMO-based protocols, e.g. 802.11n.

According to our measurement in Section 3.3, the 10ms frame length in UMTS/LTE guarantees accurate CSI estimation of BeamAdapt. However, the CSI still may become stale if there is a long interval between successive frames, e.g. 100ms (see Figure 4). To tackle this, a timing threshold $T_{th}$ is set to alarm potential expiration of the CSI to the UE. That is, if the CSI has not been updated for $T_{th}$, BeamAdapt stays with beamsteering size of one (omni directional pattern) for the next frame, even though it is not the most energy efficient pattern. The timing threshold ensures the beamsteering gain of BeamAdapt thereby the capacity requirement. To decide $T_{th}$, we leverage the recorded channel traces in Section 3.3 to get a reasonable value: Figure 10 shows the beamsteering gain under different CSI update frequencies. We adopt $T_{th}$=20ms, which allows up to 10% drop of the beamsteering gain.

### 6.3.2 Beam Pattern Adjustment

Beam pattern adjustment is implemented on top of uplink power control. Each time when the UE receives the power control command from its Node-B, i.e., the required transmit power is updated, beam pattern adjustment finds the most energy efficient beam pattern by identifying the optimal transmit power $P_{TX}$ and beamsteering size $N$. We note that a simple mapping between the required transmit power $P_O$ and the optimal beamsteering size $N$ can be established as showed in Section 4.1: $N_{opt} = \sqrt{(1 + \alpha)P_O/P_{Circuit}}$.

### 6.3.3 RF Chain Power Management

BeamAdapt adaptively changes the beamsteering size and transmit power. It allocates the total transmit power by assigning the weights to the signals. It decides the power to each individual antenna by the amplitude of the weight to its signal. Idle RF chains are powered off for power conservation. It is important to note that because the frequency synthesizer is shared by all RF chains and therefore left on, the overhead to power on/off an RF chain is very small as is already actively used in both cellular transceivers (discontinuous transmission/reception) and 802.11 transceivers (transmitter RF chain is turned off during reception).

## 7. Evaluation

In this section we thoroughly evaluate BeamAdapt with a focus on its benefits in client energy efficiency and network throughput. We employ the open-source tool Qualnet [19] for its ease of customizing protocols and support of cellular emulation. We implement BeamAdapt in the PHY layer of the UMTS protocol hierarchy. Since the beamsteering hardware is not emulated in Qualnet, we have virtually realized a beamsteering system in the UE by generating dynamic beam patterns in real time. We incorporate the power model from Section 3.2 to evaluate the energy efficiency of mobile clients.

### 7.1 Simulation Setup

We construct a scenario using close-to-reality configurations for the evaluation, shown in Figure 11. The scenario has seven Node-Bs and thirty UEs in a 4km×4km area. The Node-Bs have fixed locations, with the distance between adjacent ones set to 1.5km. While the range of each Node-B is approximately 1km, we let their coverage overlap similar to realistic cellular networks in urban areas. We also assume that the UE always uses omni directional pattern for downlink. The UEs are allowed to have mobility with random speed between zero and seventy miles per hour without being confined within the coverage of Node-B. Each UE always connects to nearest Node-B by han-doff. When a UE is out of the coverage of every Node-B, it stays idle until it moves back in the range. We set the frequency reuse factor as one as employed in current 3G cellular network; therefore each UE causes interference to all the other six Node-Bs.

We add two different applications to the UE: FTP with an unlimited-size file to transfer and constant-bit-rate (CBR) with multiple relatively small packets to deliver. FTP generates continuous traffic. CBR, on the contrary, creates intermittent traffic by the idle intervals between the small-size packets. Therefore, the FTP traffic has a higher capacity requirement than the CBR traffic.

## 7.2 Simulation Results

We evaluate the energy efficiency benefit of BeamAdapt by comparing it with the case of modern omni directional clients, in terms of UE power consumption and network throughput. We examine BeamAdapt with two, four and eight antennas, denoted as BA2, BA4 and BA8, respectively. Note that BeamAdapt with four antennas means that the client can select from one to four antennas for beamsteering. For comparison, we also demonstrate the results by Beamsteering with fixed number of antennas. That is, Beamsteering with two, four and eight antennas are denoted as BS2, BS4, and BS8, respectively.

Figure 12 shows the average power consumption of the UE for transmitting as well as the network throughput, under omni directional pattern, Beamsteering and BeamAdapt. We make several key observations. Firstly, for the FTP traffic, BA2, BA4 and BA8 saves 43%, 54% and 56% client power respectively, compared to the omni directional pattern; for the CBR traffic, the power savings are 39%, 50% and 52%. Since the FTP traffic averagely requires higher transmit power, BeamAdapt saves more power. Secondly, the energy efficiency improvement of BA8 compared to BA4 is marginal. This is due to the confined range of cellular radios and corresponding maximal transmit power limit. Therefore, one can expect that BA4 can provide enough efficiency benefit without further introducing antenna placement complexity by increasing the beamsteering size. Finally, BeamAdapt achieves approximately the same network throughput as omni-directional pattern does, which confirms that BeamAdapt always satisfies capacity requirement.

To further reveal the adaptive nature of BeamAdapt, we examine the beamsteering size statistics of four example UEs from the network, each with a different average distance, $D$, from its Node-B. Assuming BA4 is used, shows the occurrence frequency of beamsteering sizes from one to four, for each UE. We can conclude from Figure 13 that 1) under channel variation, BeamAdapt indeed behaves in an adaptive manner; 2) the most selected beamsteering size is dependent on $D$: the longer distance, the higher chance of larger beamsteering sizes. For example, UE1 with $D$=0.36km stays with omni-directional pattern ($N$=1) for more than 90% time, while UE4 with $D$=0.76km keeps using beamsteering size of four almost all the time. The small chance of $N=1$ for UE4 is due to the expiration of the timing threshold $T_{th}$. The dependency of BeamAdapt on $D$ is because longer $D$ requires larger transmit power thereby larger beamsteering size.

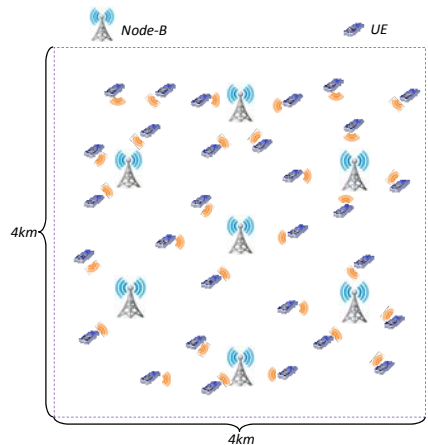

**Figure 11. Network topology and configuration for the Qualnet simulation**

## 8. Related Work

BeamAdapt is the first work that aims to enable directional communication on a mobile device using beamsteering. We next discuss related works in beamforming, directional communication and MIMO.

### 8.1 Beamforming

Existing work on beamforming, including beamsteering, does not consider circuit power and does not intend to use beamsteering on battery-constrained mobile platforms. Its focus is usually capacity and range improvement. The authors of [20, 21] show that transmit power control and beamsteering can be jointly used to improve the network capacity. The optimal transmit power and beamsteering weight vector can be iteratively identified, using the local SINR measurement. The authors assume beamsteering implemented at the base station and an omni directional antenna used at the mobile client. By considering energy efficiency and allowing beamsteering at the mobile client simultaneously with adaptive beamsteering size, we solve a much more complicated problem and address the tradeoff between network capacity and client efficiency.

### 8.2 Directional Antennas on Mobile Clients

Passive directional antenna is an efficient yet inflexible way to realize directional communication. Many have studied them for infrastructure nodes and mobile nodes that do not rotate like smartphones or Kindles, e.g. see [3, 22-27]. Most of the authors focus on MAC protocol designs to increase network spatial reuse. In contrast, our solution is

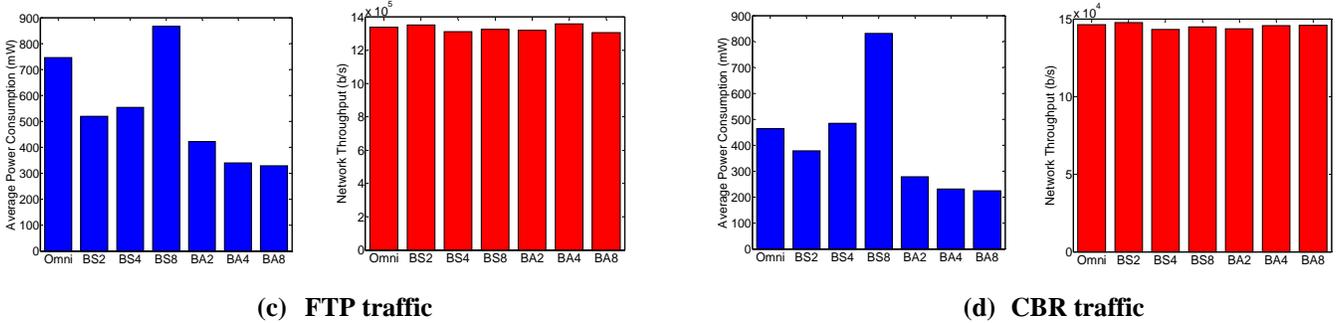

**(c) FTP traffic**  **(d) CBR traffic**

**Figure 12. Power consumption and throughput comparison between BeamAdapt and omni directional pattern**

in the PHY layer and is complementary to directional MAC designs. The flexible beam pattern can further amplify the capacity improvement of directional MACs.

Only very recently, the authors of [28, 29] demonstrated the effectiveness in improving transmission throughput and efficiency of passive directional antennas on mobile platforms that can rotate like smartphones. The solution is based on selecting one out of multiple fixed passive antennas. The key problem is to reduce the cost due to antenna assessment. In contrast, beamsteering automatically finds the right direction through CSI estimation.

### 8.3 MIMO

An alternative way to leverage the multiple RF chains and antennas is to use MIMO. By placing antennas far from each other ($>0.5\lambda$), MIMO explores diversity or spatial multiplexing gain [30]. Our previous work has shown that using the MIMO spatial multiplexing technique, one can achieve higher transceiver energy efficiency [31]. However, MIMO essentially targets at link-wise performance improvement, not being effective in reducing interference between peer links. By using beamsteering, our work goes beyond link-wise optimization and examines the tradeoff between network capacity and client efficiency.

Very recently, network MIMO [32] emerged as an effective technique to mitigate inter-cell interference and improve network capacity. While network MIMO essentially aligns and cancels interference with tight inter-link coordination, beamsteering represents an alternative way to manage interference. Instead of managing interference in a centralized way, BeamAdapt is able to accomplish it in a distributed manner to avoid the overhead of link coordination. Furthermore, beamsteering presents a unique tradeoff between client efficiency and network capacity, which BeamAdapt exploits.

### 9. Conclusions

We reported a comprehensive treatment on beamsteering on mobile devices. With both experiments and data from industry, we showed that beamsteering is not only feasible but also beneficial to mobile devices such as netbooks, eBook readers, and future smartphones in terms of energy efficiency and network capacity. Using a small network as example, we demonstrated the effectiveness of beamsteering in improving network capacity or device efficiency by focusing transmit power toward the right direction and suppressing interference to peers.

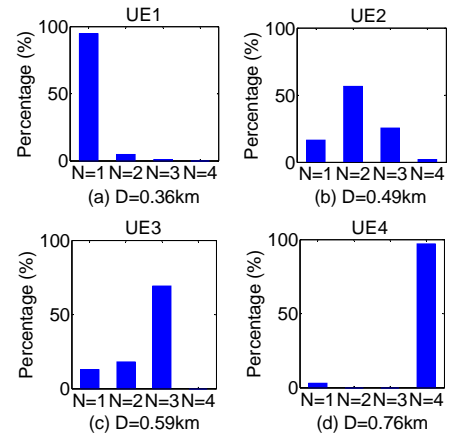

**Figure 13. Beamsteering size statistics for four UEs using BAdapt4**

We addressed the challenge of identifying the optimal operating point for a beamsteering mobile client with a distributed iterative algorithm, BeamAdapt. We demonstrated that BeamAdapt provides close-to-optimal solution with guaranteed convergence. We also showed that BeamAdapt can be efficiently implemented in modern cellular systems. Through Qualnet simulation of a large cellular network, we showed that BeamAdapt clients were able to react to mobility by promptly identifying the right number of antennas and the transmit power. Collectively they achieve the same network throughput with close to 50% lower power.

Moreover, as beamsteering essentially reduces transmit power by incurring extra circuit power, it is increasingly more efficient as semiconductor technologies continue to improve the efficiency and integration of RF circuits.

Finally, client directionality through beamsteering is a radical departure from omni directionality assumed by current mobile network paradigms. While we are able to demonstrate its benefits in client efficiency, more research at var-ious layers of the network system is required to fully realize its potential in improving the network performance.

## 10. REFERENCE

[1] V. Navda, A. P. Subramanian, K. Dhanasekaran, A. Timm-Giel, and S. Das, "MobiSteer: using steerable beam directional antenna for vehicular network access," in *Proc. Int. Conf. Mobile Systems, Applications and Services (MobiSys)*, 2007, pp. 192-205.

[2] K. Ramachandran, R. Kokku, K. Sundaresan, M. Gruteser, and S. Rangarajan, "R2D2: regulating beam shape and rate as directionality meets diversity," in *Proc. MobiSys* Poland: ACM, 2009.

[3] X. Liu, A. Sheth, M. Kaminsky, K. Papagiannaki, S. Seshan, and P. Steenkiste, "DIRC: increasing indoor wireless capacity using directional antennas," in *Proc. SIGCOMM* Barcelona, Spain: ACM, 2009.

[4] L. C. Godara, *Smart Antennas*: CRC Press, 2004.

[5] M. Blanco, R. Kokku, K. Ramachandran, S. Rangarajan, and K. Sundaresan, "On the Effectiveness of Switched Beam Antennas in Indoor Environments," in *Passive and Active Network Measurement*, 2008, pp. 122-131.

[6] D. G. Rahn, M. S. Cavin, F. F. Dai, N. H. W. Fong, R. Griffith, J. Macedo, A. D. Moore, J. W. M. Rogers, and M. Toner, "A fully integrated multiband MIMO WLAN transceiver RFIC," *IEEE Journal of Solid-State Circuits,* vol. 40, pp. 1629-1641, 2005.

[7] S. Cui, A. J. Goldsmith, and A. Bahai, "Energy-efficiency of MIMO and cooperative MIMO techniques in sensor networks," *IEEE Journal on Selected Areas in Communications,* vol. 22, pp. 1089-1098, 2004.

[8] Z. Li, W. Ni, J. Ma, M. Li, D. Ma, D. Zhao, J. Mehta, D. Hartman, X. Wang, K. K. O, and K. Chen, "A Dual-Band CMOS Transceiver for 3G TD-SCDMA," in *Solid-State Circuits Conference, 2007. ISSCC 2007. Digest of Technical Papers. IEEE International*, 2007, pp. 344-607.

[9] Z. Xu, S. Jiang, Y. Wu, H.-y. Jian, G. Chu, K. Ku, P. Wang, N. Tran, Q. Gu, M.-z. Lai, C. Chien, M. F. Chang, and R. D. Chow, "A compact dual-band direct-conversion CMOS transceiver for 802.11a/b/g WLAN," in *IEEE Int. Solid-State Circuits Conf. (ISSCC)*, 2005, pp. 98-586 Vol. 1.

[10] IEEE International Solid State Circuits Conference, *http://www.isscc.org*.

[11] IEEE Journal of Solid-State Circuits.

[12] K. Vavelidis, I. Vassiliou, T. Georgantas, A. Yamanaka, S. Kavadias, G. Kamoulakos, C. Kapnistis, Y. Kokolakis, A. Kyranas, P. Merakos, I. Bouras, S. Bouras, S. Plevridis, and N. Haralabidis, "A dual-band 5.15-5.35-GHz, 2.4-2.5-GHz 0.18um CMOS transceiver for 802.11a/b/g wireless LAN," *IEEE Journal of Solid-State Circuits,* vol. 39, pp. 1180-1184, 2004.

[13] A. Pozsgay, T. Zounes, R. Hossain, M. Boulemnakher, V. Knopik, and S. Grange, "A Fully Digital 65nm CMOS Transmitter for the 2.4-to-2.7GHz WiFi/WiMAX Bands using 5.4GHz RF DACs," in *IEEE Int. Solid-State Circuits Conf. (ISSCC)*, 2008, pp. 360-619.

[14] WARP, *http://warp.rice.edu/*, 2010.

[15] C.-K. Chau, M. Chen, and S. C. Liew, "Capacity of large-scale CSMA wireless networks," in *Proc. MobiCom* Beijing, China: ACM, 2009.

[16] G. L. Nemhauser and L. A. Wolsey, *Integer and combinatorial optimization*: Wiley-Interscience, 1988.

[17] G. J. Foschini and Z. Miljanic, "A simple distributed autonomous power control algorithm and its convergence," *IEEE Transactions on Vehicular Technology,* vol. 42, pp. 641-646, 1993.

[18] E. Dahlman, S. Parkvall, J. Skold, and P. Beming, *3G Evolution, Second Edition: HSPA and LTE for Mobile Broadband*: Academic Press, 2008.

[19] Scalable Network Technologies, *QualNet Developer: High-fidelity network evaluation software*.

[20] F. Rashid-Farrokhi, L. Tassiulas, and K. J. R. Liu, "Joint optimal power control and beamforming in wireless networks using antenna arrays," *IEEE Trans. Communications,* vol. 46, pp. 1313-1324, 1998.

[21] R. Knopp and G. Caire, "Power control and beamforming for systems with multiple transmit and receive antennas," *IEEE Trans. Wireless Communications,* vol. 1, p. 638, 2002.

[22] S. Yi, Y. Pei, and S. Kalyanaraman, "On the capacity improvement of ad hoc wireless networks using directional antennas," in *Proc. MobiHoc* Annapolis, Maryland, USA: ACM, 2003.

[23] L. Bao and J. J. Garcia-Luna-Aceves, "Transmission scheduling in ad hoc networks with directional antennas," in *Proc. MobiCom* Atlanta, Georgia, USA: ACM, 2002.

[24] K. Young-Bae, V. Shankarkumar, and N. H. Vaidya, "Medium access control protocols using directional antennas in ad hoc networks," in *INFOCOM*, 2000, pp. 13-21 vol.1.

[25] T. Korakis, G. Jakllari, and L. Tassiulas, "A MAC protocol for full exploitation of directional antennas in ad-hoc wireless networks," in *Proc. ACM Int. Sym. Mobile Ad Hoc Networking & Computing (MobiHoc)* Annapolis, Maryland, USA, 2003, pp. 98-107.

[26] R. R. Choudhury, X. Yang, R. Ramanathan, and N. H. Vaidya, "Using directional antennas for medium access control in ad hoc networks," in *Proc. ACM Int. Conf. Mobile Computing and Networking (MobiCOM)* Atlanta, GA, 2002.

[27] M. Takai, J. Martin, R. Bagrodia, and A. Ren, "Directional virtual carrier sensing for directional antennas in mobile ad hoc networks," in *Proc. MobiHoc* Lausanne, Switzerland: ACM, 2002.

[28] A. Amiri Sani, H. Dumanli, L. Zhong, and A. Sabharwal, "Power-efficient directional wireless communication on small form-factor mobile devices," in *Proc. ISLPED*: ACM/IEEE, 2010.

[29] A. Amiri Sani, L. Zhong, and A. Sabharwal, "Directional antenna diversity for mobile devices: characterizations and solutions," in *Proc. MobiCom*: ACM, 2010.

[30] A. J. Paulraj, D. A. Gore, R. U. Nabar, and H. Bolcskei, "An Overview of MIMO Communications—A Key to Gigabit Wireless," *Proc. IEEE,* vol. 92, pp. 198-218, 2004.

[31] H. Yu, L. Zhong, and A. Sabharwal, "Adaptive RF chain management for energy-efficient spatial-multiplexing MIMO transmission," in *Proc. ACM/IEEE Int. Symp. Low Power Electronics and Design* San Fancisco, CA, USA: ACM, 2009.

[32] S. Venkatesan, A. Lozano, and R. Valenzuela, "Network MIMO: Overcoming intercell interference in indoor wireless systems," in *Proc. Asilomar*, 2007, pp. 83–87.